\documentstyle[12pt]{article}

\def\ri{\ri}
\title{\bf\huge  The field equation from Newton's law of motion and absence of 
magnetic monopole} 

\author{\large  Parampreet Singh\thanks{E-mail: param@iucaa.ernet.in}
\, and Naresh Dadhich\thanks{E-mail: nkd@iucaa.ernet.in}\\
{\sl Inter-University Centre for Astronomy \&
Astrophysics,} \\
{\sl Post Bag 4, Ganeshkhind, 
Pune 411 007, India .}}

\date{}
\begin{document}
\def \n {\noindent}

\maketitle

\begin{abstract}
 By requiring the linear differential operator in Newton's law of 
motion to be self adjoint, we obtain the field equation for the linear 
theory, which is the classical electrodynamics. In the process, we are also 
led to a fundamental universal chiral relation between electric and 
magnetic monopoles which implies that the two are related. Thus there could 
just exist only one kind of charge which is conventionally called electric.

\end{abstract}

\n PACS: 03.50.De, 45.20.Dd, 03.30.+p, 14.80.Hv

\section{Introduction}

 The equations of motion for both particles and classical fields are
second order differential equations. There exist both the cases of the
equation being linear as well as non-linear. The former is the case of the
Maxwell field and motion of charged particle in it while the latter is the
case of Einstein's theory of gravitation with the geodesic motion in
curved spacetime for particles. In Einstein's theory, general
relativity (GR), the particle equation is derivable from the field
equation. Intuitively it could be simply understood as follows, since the
field is described by the curvature of spacetime and it no longer remains 
an external field, the motion of particle under gravity would naturally be
free motion relative to curved spacetime which is given by the geodesic
equation. Thus for GR, it can be said that the particle equation is  
contained in the field equation. Note that both the equations are non 
linear. 
 It raises an interesting question, does there also exist some relation 
between the equation of motion of particle and field in the case of the linear
equation? The only linear theory for a classical field is the Maxwell's 
theory of electromagnetism. That is to probe for a connection between the
equations of motion of the charged particle  and the Maxwell 
electromagnetic field. Since the equation is linear, one cannot contain
the other as was the case for Einsteinian gravitation. However, could one 
lead to the other and under what conditions? This is precisely the question 
we wish to address in this paper. 

 Let us then ask what does a linear differential equation allow us to do 
which a non linear equation does not? It allows us to construct adjoint of 
the differential operator, which can be done only for the linear operator. 
How about asking for the self adjointness of the operator? This property is 
most effectively used in the quantum theory to ensure reality of the 
eigenvalues. In the classical mechanics, it is used in identifying the 
velocity dependent potential which could be included in the 
Lagrangian$^ 1$. In general, it is not possible to incorporate 
dissipative forces in a Lagrangian (in some cases introduction of 
the Lagrange multipliers facilitates), however a particular kind 
of velocity dependent potential is permitted which is picked up by the self 
adjointness of the operator.
 The linearity of the equation implies that force on the right of the 
Newton's second law equation should involve the velocity linearly and so 
should be the case for the velocity dependent potential in the Lagrangian as 
well. That means that the potential in the Lagrangian would in addition
to a scalar have a scalar term invloving velocity linearly. 
 More elegantly, all this is ensured simply by asking that the linear 
differential operator in the Newtonian equation of motion for particle is 
self adjoint. First, it would ensure existence of Lagrangian, second, it 
would determine the force law (in the Lorentz force form) involving two
vector fields, one polar and the other axial and then their derivation 
in terms of the scalar and vector potentials given by the homogeneous 
source free set of the Maxwell equations. Note that without any reference 
to the Maxwell theory, purely from the general mechanics considerations 
follow the force law and half of the Maxwell's equations. All this would 
be true for any linear field theory and the Maxwell electrodynamics happens 
to be one such theory.
 
Following Dyson's paper$^2$ discussing Feynman's derivation of the 
homogeneous Maxwell equations, there has been spurt of activity in recent 
times in this direction. Feynman used commutation relations between 
coordinates and velocities rather than canonical momenta. It was soon 
realized that the problem is related to existence of an 
${\it Action/Lagrangian}$ for a given equation of motion. Since then a 
lot of effort has gone into building a relativistic generalization of 
Feynman's proof and its extension to non-Abelian gauge theories and to curved 
space$^ {3 - 5}$. All these attempts (with the sole exception of a recent 
paper$^6$ in which it has been shown that by incorporating magnetic 
monopoles in Feynman's formalism it is possible to derive the complete 
set of generalized Maxwell equations, though certain questions remain 
unanswered) refer only to the homogeneous set which we have simply got in by 
demanding self adjointness of the Newton's second law. Our main concern is 
thus to obtain the remaining two Maxwell equations which describe the dynamics 
of the field. This we address in the more general context of seeking 
relationship between the equations of motion for particle and field for the 
linear equation. 

 It may be noted that all these attempts involved commutation relations 
and quantum theory considerations. We would however like to stick to the 
classical mechanics and some simple general considerations. The main 
question is, could we do something imaginative to the homogeneous set 
which involves two vector fields, one each of polar and axial kind. In a 
field theory, field is produced by a source which is generally called 
charge. Without prejudice to the one or the other, let us consider the  
corresponding monopole charges for the both polar (scalar) and axial 
(pseudo scalar) fields. This will allow us to write each vector field in 
terms of a polar and axial (new) vectors by involving scalar and pseudo 
scalar charges. That is we expand our system from two to four fields and the 
two kinds of charges. Substitute this in the two homogeneous equations. We 
are led to to four equations which in parts look like the Maxwell 
equations in four vectors. It thus becomes highly under determined system.
 To proceed any further we have to contract the system back to the
two fields which we do by postulating linear relations between the two 
pairs of polar and axial vector fields. These proportionality relations
give rise to a constant which has the dimension of velocity. That is how 
an invariant speed has come up. A polar field is produced when a charge 
is at rest and an axial field is produced when it is in motion. Similarly a
pseudo charge at rest would produce axial field and polar when moving.
That means a polar/axial field could be produced by a stationary 
scalar/pseudo charge as well as by a moving pseudo/scalar charge. However 
to a test charge, it is simply a polar/axial field irrespective of its 
source. The field produced in these two different ways must be 
indistinguishable and hence there must exist a (chiral) universal 
relation between scalar and pseudo scalar charges. Then the set of equations 
in question becomes the complete set of the Maxwell's equations and the 
force law, the Lorentz force of the electrodynamics. We have thus 
obtained the equation of motion for the field corresponding to self 
adjoint Newton's law of motion. This is the complete set of the Maxwell's 
equations of classical electrodynamics. Most importantly, our method also 
leads to an important and profound result that electric and magnetic 
monopole cannot 
exist independently thereby implying that there could occur only one kind 
of charge, call it electric or magnetic. The linear theory consistent 
with the Newton's law could thus have only one kind of monopole charge.

 The paper is organized as follows. In the next Sec., we  briefly recall 
the discussion of self adjointness of the second order differential operator 
and the inverse problem in classical mechanics, which lead to the Lorentz-like
force with the homogeneous set of two equations. In Sec. III, we derive 
the intermediate set which is Galilean invariant followed by in Sec. IV 
the derivation of the entire set of the Maxwell equations and the 
fundamental relation between electric and magnetic charges. We conclude 
with a discussion of general issues and the ones to be taken up in future.

\section{Self \hspace{1mm} adjointness \hspace{1mm} and \hspace{1mm} 
 the inverse problem} 

 The inverse problem in classical mechanics deals with the demand of a
Lagrangian for a given equation of motion. 
It turns out that the sufficient condition for the existence of 
a Lagrangian 
is that the equation of motion is self adjoint . Let 
 ${\cal F}_i$ be a system of second order linear differential equations,
\begin{equation}
{\cal F}_i (t, q,\dot q,\ddot q) = 0. \hspace{8mm}   i = 1,2,...,n .
\end{equation}
If M(u) is a linear differential expression then the adjoint of M(u),
which we denote by $\bar M$(u), satisfies the Lagrange Identity
\begin{equation}
\bar v M(u) - u \bar M (\bar v) = \frac{d}{dt} Q(u, \bar v).
\end{equation}

Let M(u) be of the form 
\begin{equation}
M_i(u) = a_{ik} u^k + b_{ik} \frac{du^k}{dt} + c_{ik} \frac {d^2u^k}{dt^2}
\end{equation}
where $ a_{ik} = \frac{\partial {\cal F}_i}{\partial q^k} $,
 $ b_{ik} = \frac{\partial {\cal F}_i}{\partial \dot{q}^k} $,
and $ c_{ik} = \frac{\partial {\cal F}_i}{\partial \ddot{q}^k} $.
It  is then 
straightforward to check using the Lagrange identity that
\begin{eqnarray}
\bar M_i (\bar v) &=&  a_{ik} \bar v^k -  \frac{d}{dt}(b_{ik} \bar v^k) +  \frac {d^2}{dt^2} (c_{ik} \bar v^k) \\
Q(u, \bar v) & = & \left[ \bar v^i b_{ik} u^k + \bar v^i a_{ik} \frac{du^k}{dt} - u^k \frac{d}{dt} (\bar v ^i a_{ik}) \right ] .
\end{eqnarray}
In the case when $M(u) = \bar M(u) $ for all values of u, then M(u) is
termed as self adjoint. It turns out that necessary and sufficient conditions
for eq.(1) to be self adjoint and hence for Lagrangian to exist are $^{4,7}$


\begin{eqnarray}
\frac{\partial {\cal F} _i}{\partial \ddot q^j} & = &  \frac{\partial {\cal F} _j}{\partial \ddot q^i}  \\
\frac{\partial {\cal F} _i}{\partial \dot q^j} + \frac{\partial {\cal F} _j}{\partial \dot q^i}
& = & \frac{d}{dt} \left(\frac{\partial {\cal F} _i}{\partial \ddot q^j} +  \frac{\partial {\cal F} _j}{\partial \ddot q^i}\right) \\
\frac{\partial {\cal F} _i}{\partial q^j} - \frac{\partial {\cal F} _j}{\partial q^i} & = & 
\frac{1}{2} \frac{d}{dt} \left(\frac{\partial {\cal F} _i}{\partial \dot q^j} - \frac{\partial {\cal F} _j}{\partial \dot q^i}\right).
\end{eqnarray}   

\noindent
Eqs. (6 - 8) are known as the Helmholtz conditions and for the 
Newtonian equation of motion, we write ${\cal F}_i$ as
\begin{equation}
m \ddot q_i - F_i (t, q, \dot q) = 0
\end{equation}

\n where $F_i$ is the force experienced by a test particle.
Substitution of eq.(9) in eq.(7) yields

\begin{equation}
\frac{\partial F_i}{\partial \dot q^j} + \frac{\partial F_j}{\partial \dot q^i}
= 0 .
\end{equation}
If we substitute eq.(9) in eq.(8) we obtain
\begin{eqnarray}
\frac{\partial F_i}{\partial q^j} - \frac{\partial F_j}{\partial q^i} &=& \nonumber \frac{1}{2} \left( \frac{\partial}{\partial t} + \dot q ^k \frac{\partial}{\partial q^k} + \ddot q^k \frac{\partial}{\partial \dot q ^k}\right)\left [\frac{\partial F_i}{\partial \dot q^j} - \frac{\partial F_j}{\partial \dot q^i}\right] \\
&= & \nonumber \frac{1}{2}\left( \frac{\partial}{\partial t} + \dot q ^k \frac{\partial}{\partial q^k} \right) \left [\frac{\partial F_i}{\partial \dot q^j} - \frac{\partial F_j}{\partial \dot q^i}\right] \\
&+& \frac{1}{2} \ddot q ^k \left[\frac{\partial ^2 F_i}{\partial \dot q^j 
\partial \dot q^k} - \frac{\partial ^2 F_j}{\partial \dot q^i \partial \dot q^k}\right]. 
\end{eqnarray}
Since, the left hand side of the above equation is independent of 
accelerations, hence, for it to hold we must have
\begin{eqnarray}
\frac{\partial ^2 F_i}{\partial \dot q^j \partial \dot q^k} - \frac{\partial ^2 F_j}{\partial \dot q^i \partial \dot q^k} &=& 0 \\
\frac{\partial F_i}{\partial q^j} - \frac{\partial F_j}{\partial q^i} &=& \frac{1}{2}\left (\frac{\partial}{\partial t} + \dot q^k \frac{\partial}{\partial q^k}\right) \left[\frac{\partial F_i}{\partial \dot q^j} - \frac{\partial F_j}{\partial \dot q^i}\right] .
\end{eqnarray}
From eqs.(10 \& 12), it is easily seen that
\begin{equation}
m \ddot q_i = \lambda _i (t,q) + \xi _{ij} (t,q) \dot q^j
\end{equation}
which when substituted in eqs.(10, 12 \& 13) leads to
\begin{eqnarray}
\xi _{ij} +  \xi _{ji} &=& 0 \\
\frac{\partial \xi _{ij}}{\partial q^k} + \frac{\partial \xi _{jk}}{\partial q^i} + \frac{\partial \xi _{ki}}{\partial q^j} &=& 0 \\
\frac{\partial \xi _{ij}}{\partial t} &=& \frac{\partial \lambda _i}{\partial q^j} -  \frac{\partial \lambda _j}{\partial q^i} . 
\end{eqnarray}
Eqs.(14 - 17) are the necessary and sufficient conditions for existence of a Lagrangian 
for the Newtonian equation of motion. If we define
\begin{equation}
\lambda _i \equiv  {\cal X}_i
\end{equation}
and
\begin{equation}
\xi _{ij} \equiv  \epsilon _{ijk} {\cal Y}^k ,
\end{equation}
then eqs.(14, 16 \& 17) can be written in the vector form as
\begin{eqnarray}
\vec F &=& \vec {\cal X} + \vec v \times  \vec {\cal Y} \\
\vec \nabla \times \vec {\cal X} &=& -\frac{\partial \vec {\cal Y}}{\partial t} \\
\vec \nabla .\vec {\cal Y} &=& 0.
\end{eqnarray}
It is important to note here that $\vec {\cal X}$ and $\vec {\cal Y}$ 
are any arbitrary
fields experienced by a  test particle and eqs.(20 - 22)
will hold for ${\it any}$ Newtonian force which has self adjoint equation of
motion.
 These are the equations which were
derived by Feynman in 1948 by assuming the commutation relation between 
coordinates
and velocities.
However, these
equations can also be obtained by assuming the similar Poisson bracket 
relations $^{5,8}$. 

\section{The Galilean invariant intermediate set of equations}

 In eqs.(20 - 22), we have the Lorentz force and the homogeneous set of 
the Maxwell equations for the two fields involved. For derivation of 
the complete set of the Maxwell equations, we need only to bring the 
remaining two equations. This we shall do by first splitting the two vector 
fields into four and then recombining them. We note that $\vec F$
is a polar vector and so is $\vec {\cal X}$ while $\vec {\cal Y}$ is 
axial. We further decompose the vectors $\vec {\cal X}$ and $\vec {\cal 
Y}$ in terms of two polar $\vec {\cal E}$ \& $\vec {\cal D}$ and two 
axial $\vec {\cal B}$ \& $\vec {\cal H}$ vector fields  as follows.
 
\begin{eqnarray}
\vec {\cal X} &=& q_s \vec {\cal E} + q_p \vec {\cal H} \\
\vec {\cal Y} &=& q_s \vec {\cal B} - q_p \vec {\cal D} 
\end{eqnarray}
\n where $ q_s $ indicates a constant scalar charge and $ q_p$ the 
constant pseudo-scalar charge.

Substituting them in eqs.(20 - 22), we obtain
\begin{equation}
\vec F = q_s ( \vec {\cal E} + \vec v \times \vec {\cal B}) + q_p (\vec {\cal H} - \vec v \times \vec {\cal D}) 
\end{equation}
\begin{eqnarray}
\vec \nabla \times \vec {\cal E}  &=& -\frac{\partial \vec {\cal B}}{\partial t} \\
\vec \nabla . \vec {\cal B} &=& 0 \\
\vec \nabla \times \vec {\cal H}  &=& \frac{\partial \vec {\cal D}}{\partial t} \\
\vec \nabla . \vec {\cal D} &=& 0 . 
\end{eqnarray}
This is the intermediate set which is Maxwellian like but not quite as it 
involves four independent vector fields. It can be easily checked that 
this set is invariant under the Galilean transformation because

\begin{eqnarray}
\vec \nabla ' &=& \vec \nabla \\
\frac{\partial}{\partial t'} &=& \frac{\partial}{\partial t} + \vec V . \vec \nabla .
\end{eqnarray}

 The covariance of the force law (20) determines the following laws of 
transformations for the vector fields involved.
\begin{eqnarray}
\vec {\cal E}' &=& \vec {\cal E} + \vec V \times \vec {\cal B} \\
\vec {\cal B}' &=& \vec {\cal B} \\
\vec {\cal H}' &=& \vec {\cal H} - \vec V \times \vec {\cal D} \\
\vec {\cal D}' &=& \vec {\cal D}.
\end{eqnarray}

\section{The Maxwell equations and the fundamental relation}

 Clearly we cannot proceed further from the intermediate set (26-29) 
because it is under determined, four differential relations for 
four vector fields. In fact twice as many would be required for the 
system to be solvable. For determining a vector field both its 
divergence and curl must be given. Thus 
the integerability condition for the system requires that there
must exist the linear relations between the two polar and two axial vectors.
Secondly, field produced by a stationary scalar charge or a moving 
pseudo scalar charge must be indistinguishable for a test particle.
This would also demand a chiral relation between the charges.  
Thus we write

\begin{eqnarray}
\vec {\cal D} &=& \epsilon \, \vec {\cal E}  \\
\vec {\cal B}  &=& \mu \, \vec {\cal H}
\end{eqnarray}
and
\begin{equation}
q_p =  \sqrt{\mu / \epsilon} \hspace{0.2cm} q_s \tan \theta .
\end{equation}
Here $\epsilon$ and $\mu$ are constants and $(\mu \epsilon)^{1/2}$ has
dimensions of velocity. It is to be noted that 
in the context of charge quantization Schwinger$^{9}$ proposed a 
similar relation for dyons (particles carrying both electric and magnetic 
charge). Clearly, our context is entirely different from that of Schwinger. 
Substituting the above relations in the 
intermediate set (25-29), we obtain 
\begin{equation}
\vec F = q_s ( \vec {\cal E} + \vec v \times \vec {\cal B}) + \frac{q_p}{\mu} (\vec {\cal B} - \mu \epsilon \vec v \times \vec {\cal E}) 
\end{equation}
\begin{eqnarray}
\vec \nabla \times \vec {\cal E}  &=& -\frac{\partial \vec {\cal B}}{\partial t} \\
\vec \nabla . \vec {\cal B} &=& 0 \\
\vec \nabla \times \vec {\cal B}  &=& \mu\epsilon 
\frac{\partial \vec {\cal E}}{\partial t} \\
\vec \nabla . \vec {\cal E} &=& 0 .
\end{eqnarray}

Using eq.(38) the force law can now be rewritten as
\begin{equation}
\vec F = q_s ( \vec {\cal E} + \vec v \times \vec {\cal B}) 
+  (\mu \epsilon)^{-1/2} q_s \hspace{1mm} \tan\theta (\vec {\cal B} - \mu \epsilon \vec v \times \vec {\cal E}) 
\end{equation}
or
\begin{equation}
\vec F = q_s (\vec {\cal E} + (\mu \epsilon)^{-1/2}  \hspace{1mm} \tan\theta
\hspace{1mm} \vec {\cal B}) + q_s (\vec v \times (\vec {\cal B} - 
(\mu \epsilon)^{1/2}  \hspace{1mm} \tan\theta \hspace{1mm} \vec {\cal E}))  .
\end{equation}
We now define two fields $\vec E$ \& $\vec B$ such that 
\begin{eqnarray}
\vec E & \equiv & \vec {\cal E} + (\mu \epsilon)^{-1/2}  \hspace{1mm} \tan\theta
\hspace{1mm} \vec {\cal B} \\
\vec B & \equiv & \vec {\cal B} - (\mu \epsilon)^{1/2}  \hspace{1mm} \tan\theta \hspace{1mm} \vec {\cal E} 
\end{eqnarray}
and hence
\begin{equation}
\vec F = q_s ( \vec E + \vec v \times \vec B). 
\end{equation}
Inversely, one can rewrite eqns(46 \& 47) as
\begin{eqnarray}
\vec {\cal E} & = & \cos ^2 \theta \hspace{1mm} \vec E - (\mu \epsilon)^{-1/2} \hspace{1mm}
\cos \theta \hspace{1mm} \sin \theta \hspace{1mm} \vec B \\
\vec {\cal B} & = & \cos ^2 \theta \hspace{1mm} \vec B + (\mu \epsilon)^{1/2}
\hspace{1mm} \cos \theta \hspace{1mm} \sin \theta \hspace{1mm} \vec E .
\end{eqnarray}
Substituting these identifications of $\vec {\cal E}$ \& $\vec {\cal B}$
in eqs.(40 \& 41), we obtain
\begin{eqnarray}
 \vec \nabla \times \vec E + \frac{\partial \vec B}{\partial t}  & = &  \tan \theta \hspace{1mm} \left [(\mu \epsilon) ^{-1/2} \vec \nabla \times \vec B - (\mu \epsilon) ^{1/2} \frac{\partial \vec E}{\partial t} \right] \\
 \vec \nabla . \vec B 
 & = & - (\mu \epsilon)^{1/2}
\hspace{1mm} \tan \theta \hspace{1mm} 
\vec \nabla . \vec E \hspace{4 mm}.
\end{eqnarray}
Similarly by substituting eqs.(49 \& 50) in eqs.(42 \& 43) we obtain
\begin{eqnarray}
\vec \nabla \times \vec E + \frac{\partial \vec B}{\partial t}  & = & - \cot\theta \hspace{1mm} \left [(\mu \epsilon) ^{-1/2} \vec \nabla \times \vec B - (\mu \epsilon) ^{1/2} \frac{\partial \vec E}{\partial t} \right ] \\
 \vec \nabla . \vec B 
 & = &  (\mu \epsilon)^{1/2}
\hspace{1mm} \cot \theta \hspace{1mm} 
\vec \nabla . \vec E \hspace{4mm}.
\end{eqnarray}
The consistency of eqs.(51 - 54) demands 
\begin{eqnarray}
\vec \nabla \times \vec E  &=& -\frac{\partial \vec B}{\partial t} \\
\vec \nabla . \vec B &=& 0 \\
\vec \nabla \times \vec B  &=& \mu\epsilon \frac{\partial \vec E}{\partial t}\\
\vec \nabla . \vec E &=& 0 
\end{eqnarray}
which is the complete Maxwell set for the electric and magnetic fields, 
$\vec E$ and $\vec B$. 
\noindent
Further, note that the Lorentz
force law we have obtained could have equally been written down in terms 
of the pseudo-scalar  charge as,
\begin{equation}
\vec F = \frac{q_p}{\mu} (\vec B - \mu \epsilon \,\vec v \times \vec E)
\end{equation}
and the physics would have remained unchanged. That is,
once $q_s$ and $q_p$ are related through eq.(38), it is only a matter
of convention how does one write the force law and identifies `electric'
and `magnetic' fields. This relation means that if electric and magnetic 
charges exist, then they must be related and ultimately there is only 
one independent charge, call it electric or magnetic.
Further, since 
 $q_s$ and $q_p$ are constants, we immediately deduce from eq.(38) that
\begin{equation}
 \left(\frac{\mu}{\epsilon}\right)^{1/2} \tan \theta = K
\end{equation} 
where $K$ is a fundamental constant. Hence, we predict that the product
of the characteristic impedance $\sqrt{\mu / \epsilon}$ times $ \tan \theta $ 
is a fundamental constant.

\section{Discussion}

 We have used the self-adjointness of the differential operator in the
Newton's law and have obtained the half (homogenous) of the Maxwell equations.
In more familiar terms, alternatively this is equivalent to demanding that the force is linear in velocity and is derivable from an approporiate velocity
dependent potential. To get the other half of the equations, we resort to
the generality that there is a priori no reason to prefer one kind 
of charge over the other. Since there occur polar and axial vector 
fields, they would be produced by the corresponding scalar and pseudo scalar charges. That would lead to the two sets of the homogenous equations
in four vector fields; a pair of polar and axial vectors corresponding to the each kind of charge. This means that there are four equations for four vector
fields. It is therefore under determined and consequently unsolvable.
Finally the solvability of the system requires that the polar and axial vectors must bear a linear relation between them. Then the other set reduces to the other half of the Maxwell equations for free space. Further it would also lead to the chiral relation  between the scalar and pseudo scalar charges implying that only one kind of charge could exist. 

 Magnetic monopole was first introduced by Dirac$^ {10, 11}$ and it was 
envisioned as 
one end of an infinite string of dipoles or a solenoid. It did a 
wonderful job of quantizing electric charge even if one such entity 
existed in the whole Universe. 
The idea soon became famous and even found a rightful place in college
textbooks$^{12}$. However, if on one side Dirac's monopoles had a strong 
support and were eagerly sought by experimentalists, on the other
side it was shown that a 
theory containing them cannot be derived from an 
action principle$^ {13, 14}$. Further they led to singularity problems 
related to strings $^{15}$ and nor do they fit well in the quantum 
electrodynamics $^ {16, 17}$. Thus magnetic monopole has become an enigma, 
for it is required for the charge quantization but it could not 
successfully be accommodated in the existing theories.

The idea that a charged particle can be envisioned as carrying both
electric and magnetic charge was already there in classical physics$^{12}$,
however 
in contrary to what is prevelant in literature$^{12}$, we have convincingly
shown that both of these charges can not exist independently. One is simply
the dual of the other and what is being observed is conventionally called
`electric'. Hence, there is now little doubt on why the search for
Dirac's monopole has been futile for last 70 years.

Of 
course the question of quantization remains. For that we have to appeal 
to some quantum principle or relation. By using the Dirac 
quantization condition and the fine structure constant, it is 
straight forward to write our fundamental relation, eq.(38). In SI units the 
Dirac quantization condition is
\begin{equation}
q_s \hspace{.1cm} q_p = n \hspace{.1cm} h.
\end{equation}
where n is an integer. Using eq.(61) in eq.(60), we find the impedance $K$ as
\begin{equation}
K = n \hspace{0.1cm} \frac{h}{q_s ^2} = 2.5883... \times 10 ^4 \hspace{.1cm} n \hspace{0.2cm}ohms.
\end{equation}
Using eq.(61) in the fine structure constant relation
\begin{equation}
\alpha = \frac{q_s ^2}{4 \pi \epsilon \hbar c}
\end{equation}
we get
\begin{equation}
\frac{q_p}{\mu} = \frac{n}{2 \alpha} \hspace{1mm} q_s \hspace{.1cm} c.
\end{equation}
Defining 
\begin{equation}
\tan \theta = \frac{n}{2 \alpha}
\end{equation}
leads to our relation 
$ q_p/\mu  =  q_s \hspace{.1cm}c\hspace{1mm} \tan\theta $.
 Conversely, let us begin with the above relation and write it as $q_s\, q_p  = 
\mu \,  q_s^2 \,   c \, \tan\theta$, divide both sides by $4 \pi \epsilon \hbar c$ and choose  
$\tan\theta = n/2 \alpha$ to obtain the Dirac quantization condition 
(61).
Hence, if we take some relation from the quantum theory, then the 
charge quantization readily follows. 

A natural extension of this formalism would be for the case of particles
having internal spin. It is interesting to note that by taking the
limilt of massive wave equation for a spin-1 object, a Maxwellian
theory can be obtained$^{18,19}$. Further, it has been shown that the linearity
of wave equations is directly related to the spin for massive particles$^{20}$.
It would be worth while to probe whether all these issues can be 
reconciled with the technique proposed in this work.

The first instance of relation between the particle and field equations
was in GR where the former followed from the latter. Here we have followed
the reverse route and have obtained the latter from the former under certain general and reasonable assumptions. The question is, could for GR as well this
path be followed ? The Maxwell field equations followed by demanding the
force law to be linear in velocity and derivable from a potential. Let us 
first implement this for the relativistic particle equation, then the Maxwell
equations follow very elegantly and cogently. Now ask what field equation would
follow from a quadratic in velocity force law ? It would turn out to be 
the Einstein's equations for gravitation. This would be published soon seperately$^{21}$.
    
 Acknowledgement: We had requested several colleagues to comment on 
the first version of the paper and were glad that many of them responded 
with criticism and comments covering a wide spectrum. Not that they would all 
agree to what we have finally come up with but their response has helped 
us a great deal in achieving communicability, coherence and depth in our 
presentation. We wish to thank them all warmly and they included J. V. 
Narlikar, D. Lynden-Bell, S. Bose, S. Mukherjee, T.D. Saini, M.M. de Souza, B. Unruh and 
R.M. Wald and particularly A. Ashtekar, J. Ehlers, T. Padmanabhan and Y. Shtanov.
PS thanks Council for Scientific \& Industrial Research for grant 
number: 2 - 34/98(ii)E.U-II.

\end{document}